\documentclass[aps,10pt,prl,showpacs,twocolumn,amsmath,amssymb,superscriptaddress,footinbib]{revtex4}

\usepackage[normalem]{ulem}
\usepackage[english]{babel}
\usepackage{latexsym}
\usepackage{graphicx}
\usepackage{epsfig}
\usepackage[utf8]{inputenc}
\usepackage[T1]{fontenc}
\usepackage[usenames,dvipsnames]{color}

\def\be{\begin{equation}}
\def\ee{\end{equation}}
\def\bea{\begin{eqnarray}}
\def\eea{\end{eqnarray}}
\def\bi{\begin{itemize}}
\def\ei{\end{itemize}}
\def\bin{\begin{enumerate}}
\def\ein{\end{enumerate}}

\begin{document}

\title{Fast dynamics for atoms in optical lattices}


\author{Mateusz \L\k{a}cki}
\affiliation{
Instytut Fizyki imienia Mariana Smoluchowskiego, 
Uniwersytet Jagiello\'nski, ulica Reymonta 4, PL-30-059 Krak\'ow, Poland}

\author{Jakub Zakrzewski} 

\affiliation{
Instytut Fizyki imienia Mariana Smoluchowskiego and
Mark Kac Complex Systems Research Center, 
Uniwersytet Jagiello\'nski, ulica Reymonta 4, PL-30-059 Krak\'ow, Poland}

\date{\today}

\begin{abstract}
Cold atoms in optical lattices allow for accurate studies of  many body dynamics.
Rapid time-dependent modifications of optical lattice potentials may result in significant excitations in atomic systems.
The dynamics in such a case is frequently quite incompletely described by standard applications of tight-binding models (such as e.g. Bose-Hubbard model or its
extensions) that typically neglect the effect of the dynamics on  the transformation between the real space and 
the tight-binding basis.  We illustrate the importance of a proper quantum mechanical description using  a multi-band extended Bose-Hubbard model with time-dependent Wannier functions. We apply it to situations, directly related to experiments.
 \end{abstract}

\pacs{67.85.Hj, 03.75.Kk, 03.75.Lm}

\maketitle

Ultra-cold quantum gases in optical lattice potentials allow for precise studies of standard models known from other branches of physics  
(e.g. the condensed matter theory) as well as proposing novel situations
 with intriguing properties. The latter utilize  rich
atomic internal structures, a versatility and an extreme controllability of atomic systems \cite{Lewenstein2007,Bloch2008,Lewenstein2012}.  Many body physics often addresses stationary properties such as  phase diagrams -- cold atoms enable also a controlled study of dynamics. 
That is especially interesting in
the vicinity of quantum phase transitions \cite{Dziarmaga2010} where one of the intriguing problems is the adiabaticity  and quantitative analysis of deviation from it for  slow quenches \cite{Damski2005,Cucchietti2007,Barankov2008,Altland2009}. 

Other interesting aspects of the dynamics concern effects resulting from rapid changes of system parameters. 
The typical example of such a situation is a well known revival experiment \cite{Greiner2002,Will2010} where the sample prepared in a superfluid state is placed in the insulating environment 
by a fast increase of optical potential depth. Other possible examples include
fast quenches \cite{Klich2007,Roux2009,Roux2010,Dziarmaga2012}  and  periodic modulations of optical lattice potentials often used either for measuring
the state of the system \cite{Stoferle2004,Fallani2007} or even for modifying its effective parameters and thus its properties \cite{Eckardt2005,Eckardt2009,Zenesini2009,Struck2012,Sacha2012}. Even faster modulations were suggested to efficiently populate excited bands \cite{Sowinski2012}.

The aim of this letter is to show that for rapid modifications of the optical lattice potential (e.g. its depth) a standard application of tight-binding models is incomplete. We develop a quasi-exact multi-band theory which uses time-dependent Wannier functions and show its applicability on few chosen model examples. 

For weak optical potentials, in the deep superfluid regime, the standard approach is to use a Gross-Pitaevski mean-field approach \cite{Gross1961,pit}. In deeper lattices, 
the depletion of the condensate becomes significant and another approach is necessary. The seminal work \cite{Jaksch1998} uses a tight-binding approach mapping the real space system onto
the lattice, resulting in a famous Hubbard model for fermions \cite{Hubbard1963} or the so-called Bose-Hubbard model for bosons.  One may consider 3D cubic lattices realized by three orthogonally polarized laser standing waves.
Often the reduced, two- (2D) and one-dimensional (1D)
geometries are interesting \cite{Stoferle2004,Fallani2007} for which very deep lattices in remaining directions cut the atomic sample into 2D slices or 1D tubes (with the confined degree(s) of freedom 
effectively described by the harmonic oscillator ground state). Explicitly, for the simplest quasi-1D situation
\cite{remark0} 
  $V_{1}(\vec r)=s \sin^2(k x) + \frac{m}{2}[ \omega^2(y^2 + z^2)+\omega_x^2 x^2],$
where  $\omega_x^2 x^2$ is an additional traping potential.
Parameters $s, \omega,\omega_x$ are tunable in the experiment.

Cold interacting Bose gas  described by a second quantized Hamiltonian: 
\bea
 {\cal H}&= &\int \textrm{d}^3\vec r \Psi^\dagger(\vec r)\left(-\frac{\hbar^2}{2 m } \nabla^2+V_{1}(\vec r)\right)\Psi(\vec r)  \nonumber\\
&+&\frac{1}{2}\int\textrm{d}^3\vec r \textrm{d}^3\vec r'  \Psi^\dagger(\vec r) \Psi^\dagger(\vec r')V(\vec r, \vec r') \Psi(\vec r) \Psi(\vec r'), 
\label{eq:Hx}
\eea
where $V(\vec r,\vec r')$ is  an isotropic short-range pseudopotential modelling s-wave  interactions \cite{Bloch2008}  
\be 
V(\vec r,\vec r')=\frac{4\pi \hbar^2 a}{m}\delta(\vec r-\vec r')\frac{\partial}{\partial |\vec r-\vec r'| }|\vec r-\vec r'| 
\ee
 with $a $ being the scattering length. 

The field, $\Psi(\vec r)$, is expanded in basis functions, $W^\alpha_i(\vec r,s)=w^{\alpha}_i(x,s)H(y)H(z)$  [built as a product of Wannier functions in the direction of the lattice with  harmonic oscillator functions in transverse direction: $H(z)=({\kappa}/\pi)^{1/4}\exp(-{\kappa}z^2/2)$], $\kappa=m\omega/\hbar$:
\be
 \Psi(\vec r)=\sum_{i,\alpha} a_i^\alpha W^\alpha_i(\vec r,s),
\label{eq:Psi}
\ee
 $i$ numbers the sites and $\alpha$ Bloch bands of the lattice.

Performing the integrations in (\ref{eq:Hx}) using orthogonality of Wannier functions the extended Bose -Hubbard (EBH) Hamiltonian is obtained
\begin{eqnarray}
&&\mathcal{H} = -\sum\limits_{i\ne j,\alpha}J_{i-j}^{\alpha} ((a_i^\alpha)^\dagger a_{j}^\alpha  + h.c.)+ \sum_{i,\alpha} E_i^\alpha n_i^\alpha+ \nonumber\\
&&\frac{1}{2}\!\!\sum\limits_{\alpha,\beta,\gamma,\delta}\sum\limits_{ijkl} U_{ijkl}^{\alpha\beta\gamma\delta} ({a^{\alpha}_i})^\dagger ({a^{\beta}_j})^\dagger a^{\gamma}_k a^{\delta}_l
\label{eqn:ebh},
\end{eqnarray}
with the $J$-terms describing tunnelings between sites while the $U$-terms 2-body collisions. Explicitly,
$ J_{i-j}^\alpha=\int W^\alpha_{i}(\vec r) \left(-\frac{\hbar^2}{2 m }\nabla^2 + V_1(\vec r) \right) W^\alpha_{j}(\vec r ) \textrm{d}^3\vec r,$ while  $U_{ijkl}^{\alpha\beta\gamma\delta} = \int \textrm{d}^3\vec r  \textrm{d}^3\vec r' W_i^\alpha(\vec r)W_j^\beta(\vec r')V(\vec r, \vec r')W_k^\gamma(\vec r)W_l^\delta(\vec r').$ 
$E_i^\alpha=J_{0i}^\alpha$ depend on $i$ via the trapping potential and 
may be expressed as $E_i^\alpha=E^\alpha +K(i-i_0)^2$ with $K$ being the curvature of the trap.

The EBH Hamiltonian, (\ref{eqn:ebh}) requires simplifications to be of a practical use.
For sufficiently deep lattices (say $s\ge3$) we may restrict the tunneling to nearest neighbors only (see \cite{Trotzky2012} for a shallow lattice case when next nearest neighbor tunnelings also play a role). Consistently, for interactions, we include  terms such that $i=j=k=l$ or $i=j=k$, and $l$ being nearest neighbor of $i$ up to a permutation (the so called density dependent tunnelings \cite{Dutta2011,Luhmann2012,Lacki2012a} are taken into account).  From now on by EBH we shall denote the Hamiltonian (\ref{eqn:ebh})
with the finite, low number of bands: $\alpha=0,1,..,B$ 
(so large energies where details of the real interaction potential become important are avoided).
 The corresponding Wannier functions are smooth and the action of the pseudo-potential is equivalent to a standard contact term $V(\vec r,\vec r')=\frac{4\pi \hbar^2 a}{m}\delta(\vec r-\vec r')=g\delta(\vec r-\vec r')$.
Restricting $\alpha$ to the lowest band only ($\alpha=0$) and taking solely $i=j=k=l$ on-site interactions gives the standard Bose-Hubbard (BH) model \cite{Jaksch1998}. In the following we adopt the recoil energy 
$E_R=\frac{k^2 \hbar^2}{2m}, k=\frac{2\pi}{\lambda} $ as an energy unit ($\lambda$ is a wave of the laser). We take $k^{-1}=\lambda/2\pi$ as the unit of length.
 
It is vital to note that Wannier functions depend on the lattice parameters, in particular $s.$
While EBH genuinely describes the dynamics in such a lattice, we run into problems when, e.g. the lattice depth $s$ varies in time. There are two options:  either we keep the basis fixed in time determining it once at a given, say initial, $s=s_0$ value or we make the basis {\it time dependent} so that Wannier functions change with time [i.e. we use $s(t)$ instead of $s_0$ in Eq.~(\ref{eq:Psi})]. The former, while conceptually simpler, leads to difficulties: once $s$ in the Hamiltonian (\ref{eq:Hx}) is different than $s_0$ chosen for Wannier functions, the resulting Hamiltonian is no longer in the form of EBH as defined in (\ref{eqn:ebh}) - in particular tunneling-type terms appear between different bands. Therefore most of the authors use the latter approach (at least implicitly) using the EBH (or BH just for the lowest band). Then changes of, e.g., lattice depth $s(t)$
are just translated into changes of Hamiltonian parameters $J_{i-j}^\alpha$ and $U_{ijkl}^{\alpha\beta\gamma\delta}$ evaluated for $s_0=s(t)$ (see e.g. \cite{Greiner2002,Zwerger2003,Sowinski2012,Stoferle2004,Kollath2006}). 

\begin{figure}[ht]
\centering
\includegraphics[clip,width=8.5cm]{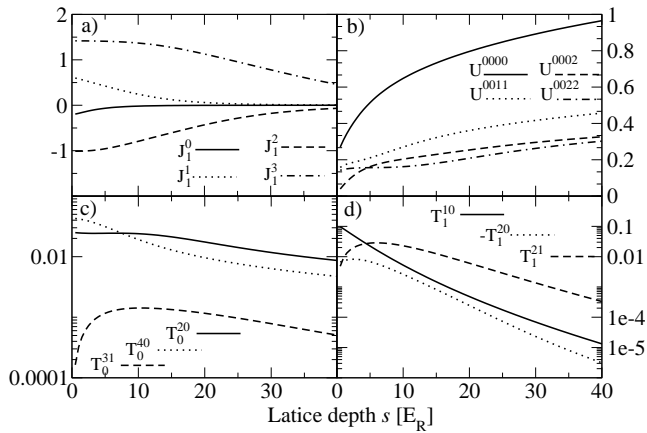}
\caption{ (color online) Relevance of different transition amplitudes:  (a) -- nearest neighbor tunnelings $J_1^\alpha$ for different Bloch bands    (b) -- interaction integrals for $g=1, \kappa=2\pi$;  the term $U_{iiii}^{0000}$ term present in the Bose-Hubbard Hamiltonian is compared with interaction terms involving excited bands.Panels (c) and (d) show additional amplitudes $T^{\alpha\beta}_{i-j}$ [see Eq.~ (\ref{eqn:wdot})] coming from the time-dependence of 
Wannier functions 
[i.e. ${\cal W}$ term (\ref{eq:trans})].
}
\label{fig:valWdot}
\end{figure}

Such an approach {\it neglects} the time-dependence of Wannier functions. 
The situation is similar to a basic textbook unitary transformation case. Recall that
if $\psi(t)={\cal U}(t)\chi(t)$ and the evolution of $\chi$ is governed by the Hamiltonian $H$ then the proper Hamiltonian for time evolution of $\psi$ is  ${\cal U} H {\cal U}^\dagger(t)+i\hbar (d/dt {\cal U}(t)) {\cal U}^\dagger(t)$. Using time dependent Wannier functions is equivalent to performing a similar transformation on our system. A straightforward calculation \cite{remark0} yields the proper 
Hamiltonian ${\cal H_W}$ in the instantaneous Wannier basis in the form
\be 
{\cal H_W}={\cal H}+{\cal W}={\cal H} +i\hbar  \frac{d}{dt} ({\cal U}(t)) {\cal U}^\dagger(t).
\label{eq:trans}
\ee
Specifying further on that we are interested in changes of lattice depth in time we may write $d/dt {\cal U}(t) = \partial_s ({\cal U}) ds/dt$ and ${\cal W}={\cal T}ds/dt$. Then we obtain \cite{remark0}

\begin{eqnarray}
\mathcal{T}&=&-i\!\!\sum\limits_{i,j,\alpha,\beta}  T_{i-j}^{\alpha\beta}(s) (a_i^{\alpha})^\dagger a_j^{\beta}\nonumber\\
T_{i-j}^{\alpha \beta }(s)&=& \int w_i^{\alpha} (x,s) \frac{d}{ds}{w}_j^{\beta}(x,s) \textrm{d}x.
\label{eqn:wdot}
\end{eqnarray}
Transition integrals $T_{i-j}^{\alpha\beta}(s)$ obey relations:  $\forall i,j,\alpha: T_{i-j}^{\alpha\alpha} \equiv 0$, $T_{i-j}^{\alpha\beta}=(-1)^{\alpha+\beta}T_{j-i}^{\alpha\beta}=-T_{j-i}^{\beta \alpha}$, $T_{0}^{\alpha\beta}=0$ for $\alpha-\beta$ odd. In particular, ${\cal T}$-term correction to a single band model (e.g. a standard BH) is zero. Thus the influence of the ${\cal W}$ term is expected only when coupling to higher bands is appreciable.
The ${\cal T}$ term contains both on- and off-site terms. Practically, most significant coupling occurs on-site between bands $\alpha=0$ and $\alpha=2$. $T_{d}^{\alpha\beta}(s)$ decrease rapidly as $|d|$ grows. In Fig.~\ref{fig:valWdot} most prominent of these parameters are shown as a function of $s.$ In numerical examples below terms up to nearest neighbors are taken only.

As mentioned above the ${\cal W}$ term is usually omitted in  numerical simulations. Its  importance depends on the value of $\frac{ds}{dt}$. Rapid changes of the lattice in time are necessary for the effects of ${\cal W}$  to be appreciable.
Thus for slow (e.g. 100ms) quasi-adiabatic quenches from shallow to deep optical lattices leading to Mott insulator phase formations \cite{Greiner2002,Stoferle2004,Fallani2007,Zakrzewski2009}, the ${\cal W}$ term may be ignored safely.

The situation becomes different for fast changes of $s$. We shall
consider the  influence of ${\cal W}$ term in model situations. We shall discuss first a simple model of a linear quench. Then we briefly mention the famous revival experiment \cite{Will2010}. Finally we analyze the recent proposition for efficient higher bands excitations \cite{Sowinski2012}.

\begin{figure}[ht]
\centering
\includegraphics[clip,width=8.5cm]{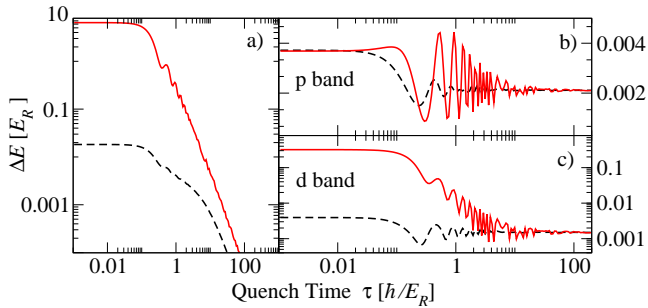}
\caption{ (color online) $\Delta E$,  the energy gain (i.e. the excess energy over the corresponding ground state) during a linear quench of a model 1D system from $s_0=12 E_R$ to $s_1=40 E_R,$. For the adiabatic process $\Delta E=0$.  Red (black, dashed) lines correspond to the simulation with (without)
the ${\cal W}$ term (\ref{eq:trans}) in the time-dependent Hamiltonian.  Panels b) and c) show time variation of averages of band occupation operators $\langle \hat{n}_1  \rangle,$ $\langle \hat{n}_2  \rangle.$    
}
\label{fig:eneQuencz}
\end{figure}

{\it A linear quench.} It is realized assuming  $s(t)= s_1 t/\tau + (1-t/\tau) s_0,$ where $\tau$ is the duration of the quench. Consider   $N=5$ $^{87}\textrm{Rb}$ atoms placed in a 1D lattice of length $L=4$  under periodic boundary conditions (PBC). The exact diagonalization gives the ground state  at $s_0=12$ -- the initial state. The quench is performed up to $s_1=40$, for different values of the quench time, $\tau$.
 We find that  (see Figure \ref{fig:eneQuencz}) as soon as $\tau<\hbar/E_R$ the excitation energy becomes significantly larger in the presence of  the ${\cal W}$ term  than without it. That reflects a significant differences in the occupation of the second excited band. Thus a simple treatment of higher bands via the EBH model is insufficient to explain the dynamics; time-variation of Wannier functions has to be taken into account.

In the limit $\tau\to 0$ the quench becomes instantaneous, the evolution does not change the wavefunction.
Yet the Wanner function basis changes from $\mathcal{B}(s_{0})$ to $\mathcal{B}(s_{1})$ and so does the field operator (\ref{eq:Psi}) representation
in the Wannier function basis. The basis change is realized via ${\cal U}(s_1){\cal U}^\dagger(s_0)$ operation obtained directly from Eq.(\ref{eqn:wdot}).

In the revival experiment \cite{Will2010}  a rapid quench is realized for bosons in the optical lattice by a rapid increase of the lattice depth.
The authors were of course aware of the fact that too fast a quench would have populated excited bands thus they chose the duration of the quench, $\tau=50 \mu s$, sufficiently long so that 
the higher bands excitations were negligible. As it turns out already for 
$\tau=20\mu s$ effects due to Wannier function dynamics are appreciable
\cite{remark0}. 

{\it  Higher bands excitations.}  Recently Sowi\'nski suggested \cite{Sowinski2012} to use periodic modulations of the lattice depth, say in the $x$-direction,
\be
 s_x(t)= s^0_x + s_m \sin(\omega t),
\label{st}\ee
 for excited orbital quantum state preparation. 
The large energy gap between Bloch bands requires  high frequency driving to couple different Bloch band states \cite{Sowinski2012}. This translates into significant values of $ds/dt$ and contributions from the ${\cal W}$ term in the dynamics may be significant - these terms were not taken into account in \cite{Sowinski2012}. 

\begin{figure}[ht]
\centering
\includegraphics[clip,width=8.5cm]{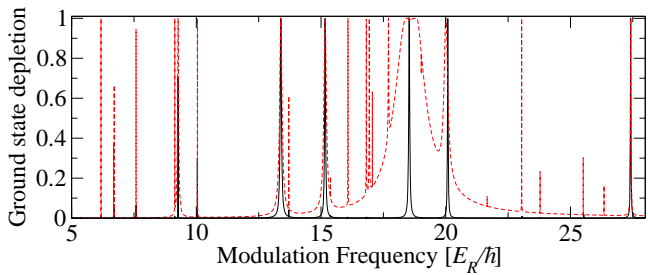}
\caption{ (color online) Excitation via modulation of the lattice depth with and without $\cal{W}$ term, Eq.~(\ref{eq:trans}). The depletion function during the first 10ms, without (black, solid) and with the ${\cal W}$ term (red, dashed). The broadening of the peak around $\omega=18.5$ is a power broadening effect, see discussion in the text.
}
\label{fig:Sow2606}
\end{figure}

To see how the important the ${\cal W}$ part is we have recalculated the numerical simulation \cite{Sowinski2012} using (\ref{eqn:ebh}) with and without the ${\cal W}$ term, Eq.~(\ref{eq:trans}-\ref{eqn:wdot}). The studied system is a 2D lattice with the tight harmonic confinement in the third direction \cite{remark0}. The system is assumed to be in a deep Mott insulator regime $(s_x=32,s_y=20,\kappa=8).$  Due to a deep lattice potential, the inter-site hopping can be neglected, and the whole system decouples into independent 2D sites with the (assumed) integer filling $\nu=2.$ We have prepared the system in the ground state $|\psi(0)\rangle$ with energy $E_0.$ Following \cite{Sowinski2012} we restrict the numerical simulation to first three bands (while this may not be sufficient for a simulation of a real situation since ${\cal W}$ terms efficiently populate higher bands we consider the {\it same} model as in \cite{Sowinski2012} to isolate the influence of Wannier functions' dynamics). We have performed the numerical evolution of the system for time corresponding to $10 \textrm{ms}$  with varying frequencies $\omega_x.$ As in \cite{Sowinski2012} we measure the maximal ground state depletion:  $\delta(\omega_x) = 1-\sup_{t\in [0,T]}  |\langle \psi_{\omega_x}(t) | \psi(0)\rangle| $  as a function of the driving frequency $\omega_x.$ We find that the presence of ${\cal W}$ term  changes significantly the depletion function in the frequency range considered (compare  Fig.\ref{fig:Sow2606}). The ${\cal W}$ term leads to several additional excitations accompanied also by
 broadening and  shifting the excitation peaks obtained without the ${\cal W}$ term. 

A key result of \cite{Sowinski2012}  is a possibility of efficient population of higher Bloch bands via Rabi-like oscillations. We have found that including the ${\cal W}$ term in the analysis makes the process much faster and efficient. The oscillation period is decreased usually several times with similar excitation efficiency. Therefore, while confirming the possibility of direct resonant transfer of population
to excited bands by lattice depth modulation, our analysis suggests that
taking the time variation of Wannier functions into account is crucial for
controlling the process and for selective excitation of desired bands.  

\begin{figure}[ht]
\centering
\includegraphics[clip,width=3.1cm,angle=-90]{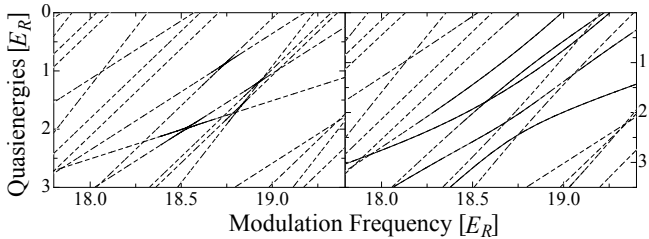}
\caption{Floquet spectrum without ${\cal W}$ contribution (left) and with  with ${\cal W}$ term, Eq.~(\ref{eq:trans}) (right panel). Broad avoided crossings for the latter are due to the strength of terms omitted within BH description as well as the influence of higher harmonics - see text for discussion. The region of avoided crossings (large curvatures) is highlighted with solid lines. 
}
\label{fig:floq}
\end{figure}
The presence of the ${\cal W}$ term in the Hamiltonian, (\ref{eq:trans}), may be also analyzed
using the Floquet approach \cite{Eckardt2005}. Exemplary spectra inspecting the broad structure around $\omega=18.5$ in the depletion function are shown in Fig.~\ref{fig:floq}. The spectrum without ${\cal W}$ contribution shows a single isolated avoided crossing indicating a simple resonance (corresponding to the isolated peak in Fig.~\ref{fig:Sow2606})
in a contrast to broad structure of avoided crossings for the quasi-exact evolution. This structure correlates well with the broadened peaks observed in the depletion function.

To understand this striking difference it is enough to consider the magnitude of different oscillatory terms in the Hamiltonian. For a deep lattice, when the tunneling is negligible and a single site is considered only \cite{Sowinski2012} the driving comes from the time dependence interaction term as well as from ${\cal W}$ term. The dominant interaction term $U^{0000}$ may be expanded as [compare (\ref{st})] 
\be 
U^{0000}(s)\approx U^{0000}(s^0_x) + \frac{\partial U^{0000}}{\partial s_x}s_m \sin \omega t \approx 1.45+0.050  \sin \omega t
\label{ut}
\ee
for $s^0_x=32$ and $s_m=4$ (in recoil units). The dominant intrasite hopping term coming from the ${\cal W}$ term 
\bea
T^{20}_0(s)\frac{ds_x}{dt}&\approx& \left(T^{20}_0(s^0_x)+ \frac{\partial T^{20}_0}{\partial s_x}s_m \sin \omega t\right)s_m\omega\cos\omega t \cr\nonumber
&\approx& 0.79 \cos\omega t-0.026\sin 2\omega t
\label{tt}
\eea
for $\omega=18.5$. Thus the driving term coming from ${\cal W}$ contribution is an order of magnitude stronger than driving induced by a modulation of interactions! The presence of an ``in phase'' and ``in quadrature'' driving breaks the time-reversal invariance \cite{Sacha1999} that strongly affects the structure of avoided crossings \cite{Zakrzewski1991,Zakrzewski1993}. Moreover  ${\cal W}$ term brings strong second harmonic which additionally strongly modifies Floquet spectrum as well as the dynamics. For parameters of \cite{Sowinski2012} the 
 ${\cal W}$  term dominates the dynamics.
 
Let us mention also that the importance of the ${\cal W}$ term in the evolution appears also for sufficiently fast oscillatory movement of the lattice sites with a fixed lattice depth \cite{Eckardt2005,Arimondo2012}. The selection rules for excitations  are then modified with respect to the simple, translationally invariant, symmetric situation discussed here \cite{remark0}. 
Other possible applications of the presented formalism may be thought of due to the generality of the arguments leading to Eq.~(\ref{eq:trans}).
In particular, with appropriate modifications it can be formulated for fermions as well.

In conclusion we have shown that great care has to be taken when using discretized tight-binding approximations for atomic systems in optical lattices in the presence of fast time variations of optical lattice parameters. A transformation to Wannier functions basis involves time dependent terms reflecting the time dependence of Wannier functions themselves. These terms modify the tight-binding Hamiltonian and are responsible for appreciable and directly experimentally measurable effects whenever the change of lattice parameters is sufficiently fast.  

Recently a variational approach for studying non-equilibrium dynamics has been developed using time-dependent  Wannier basis \cite{Sakmann2011}. The variational ansatz assumes a {\it single} time dependent Wannier function per site (thus all particles at a given site must be in the same mode). While this approach seems promising in certain applications, it fails to treat real transitions to excited bands whenever there are two or more particles per site (e.g. it cannot represent properly the entangled two particle state with one of them in the lowest, the other in some higher band). A detailed comparison of both approaches with the possible generalizations will be performed elsewhere.

We acknowledge discussions with Dominique Delande and Krzysztof Sacha on the final form of this work. Projects of M.\L. and J.Z. have been financed by Polish National  Science Center under contract DEC-2011/01/N/ST2/02549    and DEC-2012/04/A/ST2/00088 respectively.


\bibliographystyle{apsrev}

\section{Supplementary material to: 
Fast dynamics for atoms in optical lattices}


{\bf Mateusz \L\k{a}cki and
Jakub Zakrzewski}

Abstract:

We provide a detailed derivation of the tight binding Hamiltonian which takes into account time-dependence of Wannier functions. We provide expressions valid for different  dimensionalities of the optical lattice potentials.



\subsection{ Multiband Bose-Hubbard Hamiltonian}

To make this material self-contained we repeat below some of the formulae
from the letter extending it at the same time by pedagogical comments.

Cold interacting Bose gas in an external potential $V(\vec r)$ may be described by a following second quantized Hamiltonian: 
\bea
 {\cal H}&= &\int \textrm{d}^3\vec r \Psi^\dagger(\vec r)\left(-\frac{\hbar^2}{2 m } \nabla^2+V(\vec r)\right)\Psi(\vec r)  \nonumber\\
&+&\frac{1}{2}\int\textrm{d}^3\vec r \textrm{d}^3\vec r'  \Psi^\dagger(\vec r) \Psi^\dagger(\vec r')V(\vec r, \vec r') \Psi(\vec r) \Psi(\vec r'), 
\label{eq:Hxa}
\eea
where $V(\vec r,\vec r')$ is an interaction potential between bosons. For a very low temperature, the $s$-wave scattering dominates and the real potential may be represented as  an isotropic short-range pseudopotential  \cite{Bloch2008}  
\be 
V(\vec r,\vec r')=\frac{4\pi \hbar^2 a}{m}\delta(\vec r-\vec r')\frac{\partial}{\partial |\vec r-\vec r'| }|\vec r-\vec r'| 
\label{pot1}
\ee
 with $a $ being the $s$-wave scattering length. 

We shall consider an external potential corresponding to a cubic optical lattice \cite{trap} (generalizations to other lattice configurations are quite straightforward):
\be
V(\vec r)= \sum_{t=x,y,z} s_t \sin^2( k \vec e_t\cdot \vec r)
\ee
where $s_t$ is the depth of the periodic standing wave potential in the $t$ -direction, $\vec e_t$ - versor in the corresponding direction, while
$k=2\pi/\lambda$ the wavevector for light with wavelength $\lambda$.  Consider $x$ direction only and
 recall that Wannier functions, $w_i^\alpha(x,s)$, localized  at lattice sites, are linear combinations of Bloch functions $\phi^\alpha_p$, eigenfunctions of the single particle problem with potential $s\sin^2(kx)$  and quasimomentum {$p\in[-k,k)$} \cite{Kohn1959} :

{
\be
w_i^\alpha(x,s)=\sqrt{\frac{1}{4\pi k}}\int\limits_{-k}^{k} \phi_p^\alpha(x) dp.
\ee
}

Now the field operator, $ \Psi(\vec r)$, is expanded in terms of three-dimensional (3D) Wannier functions $W_i^\alpha(\vec r, \mathbf{s})$ (for the potential depths $\mathbf{s}=(s_x,s_y,s_z)$) that are products of the corresponding one-dimensional Wannier functions $w_i^\alpha(s_t, t)$ 
\be
 \Psi(\vec r)=\sum_{i,\mathbf{\alpha}} a_i^\alpha W^\alpha_i(\vec r,s),
\label{eq:Psi}
\ee 
with 
\be
W_i^{\mathbf\alpha}(\vec s, \vec r)=w_i^{\alpha_x}(s^x,x) w_i^{\alpha_y}(s_y,y) w_i^{\alpha_z}(s_z,z).
\label{eqn:bf3}
\ee
In the expressions above  the multiindex $\mathbf{\alpha}=(\alpha_x,\alpha_y,\alpha_z)$ numbers different Bloch bands of the lattice while
$a_i^\alpha$ is an annihilation operator for bosons at the site $i$ and in the $\alpha$ band. Index $i$ numbering the sites is, also really a multiindex $i=(i_x,i_y,i_z)$.

 Using the orthogonality of Wannier functions and  performing  integrations in (\ref{eq:Hxa})
the extended Bose -Hubbard (EBH) Hamiltonian is obtained

\begin{eqnarray}
&&\mathcal{H} = -\sum\limits_{t={x,y,z}}\sum\limits_{ i \stackrel{t}{\to} j}\sum\limits_{\mathbf\alpha} J^{\alpha_t}_{i-j}(s_t) (a_i^{\mathbf\alpha})^\dagger a_j^{\mathbf\alpha}  +  \nonumber\\
 &&+\sum\limits_{\mathbf\alpha} E_{\mathbf\alpha} n_{\mathbf\alpha}+ \frac{1}{2}\sum\limits_{\mathbf\alpha\beta\gamma\delta}\sum\limits_{ijkl} U_{ijkl}^{\mathbf\alpha\beta\gamma\delta} ({a^{\mathbf\alpha}_i})^\dagger ({a^{\mathbf\beta}_j})^\dagger a^{\mathbf\gamma}_k a^{\mathbf\delta}_l
\label{eqn:ebha}.
\end{eqnarray}

The notation ${ i \stackrel{t}{\to} j},$ following \cite{Sowinski2012} in
the sum indicates summation over all  sites $i$ and $j$ shifted in $t$ direction by distance $i-j$.
The $J$ terms describe tunneling between sites 
\be
J_{i-j}^{\alpha_t}=\int W^\alpha_{i}(\vec r) \left(-\frac{\hbar^2}{2 m }\nabla^2 + V(\vec r) \right) W^\alpha_{j}(\vec r ) \textrm{d}^3\vec r,
\ee
while $U$ terms 2-body collisions
\be
U_{ijkl}^{\alpha\beta\gamma\delta} = \int \textrm{d}^3\vec r  \textrm{d}^3\vec r' W_i^\alpha(\vec r)W_j^\beta(\vec r')V(\vec r, \vec r')W_k^\gamma(\vec r)W_l^\delta(\vec r').
\ee

 The on-site energies 
$E_i^\alpha=J_{0}^\alpha$ do not depend on site for translationally invariant systems (we leave this dependence to signify that one may  incorporate easily additional slowly varying inhomogeneous term).  

Formally this representation of the Hamiltonian  is {\it exact} (formally in a sense that the convergence of infinite series expansions requires a missing proof). The approximations emerge when we limit the number of bands as well as put restrictions on the Hamiltonian parameters. In particular, restricting the 
expansion to the lowest band only, assuming that the tunneling amplitudes are non-zero to nearest neigbors while interactions are entirely on-site, we recover the standard Bose-Hubbard Hamiltonian  \cite{Jaksch1998}. Since we will be interested in processes populating excited bands we shall take them into account, limiting ourselves to several of them in each direction.

As long as the number of bands included is limited to a {finite} number, any function {built} as a linear combination of Wannier functions will be smooth.
Then a pseudopotential, Eq.(\ref{pot1}), may be replaced by a  Fermi delta-potential
\be V(\vec r,\vec r')=\frac{4\pi \hbar^2 a}{m}\delta(\vec r-\vec r').
\label{pot2}
\ee

From now on we shall assume a convenient units in which energy is expressed in recoil units (with recoil energy $E_R=\frac{\hbar^2k^2}{2m}, k=\frac{2\pi}{\lambda}$), and $\frac{\lambda}{2\pi}=1/k$ is the unit of length.
The potential, (\ref{pot2}) is then 
\be V(\vec r,\vec r')=g\delta(\vec r-\vec r'),
\label{pot3}
\ee
with dimensionless coupling constant $g=8\pi a k$. The interaction integrals 
separate into a product corresponding to each coordinate taking the form
\bea
U_{ijkl}^{\alpha\beta\gamma\delta} &=& g \prod_{t=x,y,z} u_{ijkl}^{\alpha_t\beta_t\gamma_t\delta_t}\nonumber\\
&=&g \prod \int \textrm{d}t  w_i^{\alpha_t}(t)w_j^{\beta_t}(t)w_k^{\gamma_t}(t)w_l^{\delta_t}(t).
\eea

Frequently deep optical lattices are used to to separate atomic cloud into parts, a single {retroreflected} beam in one direction cuts a cloud into 2D slices, {two} such perpendicular standing waves produce a set of very weakly coupled tubes. Separating different energy scales one may in those situations assume, that in the tightly confined direction (directions for tubes) the system remains in 
the lowest band and neglect the tunnelings between slices (tubes). This has to be done with care, virtual effect of high lying excited bands may be not negligible \cite{Dutta2011,Luhmann2012,Lacki2012a}. 
 In this limit one may often approximate the Wannier function in that lowest band, say in $z$ direction by a ground state
of an appropriate harmonic oscillator, i.e $w_i^{0_z} \approx H(z)=({\kappa}/\pi)^{1/4}\exp(-{\kappa}z^2/2)$, where dimensionless $\kappa=\hbar\omega_z/2E_R=\sqrt{s_z/E_R}$ where $s_z$ corresponds to the lattice depth in that direction. Thus the basis functions for 2D system with
the effective potential of the form
\be  
V_{2}(\vec r)=s_x \sin^2(k x) + s_y\sin^2(k y) +  \frac{m}{2} \omega_z^2 z^2,
\ee
are 
\be
 W^\alpha_i(\vec r,s)=w^{\alpha_x}_i(x,s_x)w^{\alpha_y}_i(y,s_y)H(z).
\label{eqn:bf2}
\ee
For 1D tube, the corresponding formulae read
\be
V_{1}(\vec r)=s \sin^2(k x) + \frac{m}{2} \omega^2(y^2 + z^2),
\ee and
\be 
W^\alpha_i(\vec r,s)=w^{\alpha}_i(x,s)H(y)H(z).
\label{eqn:bf1}
\ee

Observe that in the reduced geometries tunneling takes place along the periodic lattice potential only. This simplifies the notation, in particular in 1D geometry, the index $\alpha$ numbering the bands is {simply an integer}, $\alpha=0,1,..,B$,
the tunneling is unidimensional and Hamiltonian (4) of the paper is obtained as
an extended Bose Hubbard model.

\subsection{Contribution of the time dependence of Wannier functions to the tight-binding Hamiltonian} 
Suppose that we are interested in the dynamical problem with time dependent lattice depth $s=s(t), s(0)=s_0$.  The initial wavefunction is expressed in Wannier basis $\{W_i^\alpha(\cdot, s_0)\}.$ Similarly the EBH Hamiltonian is obtained for that value. As discussed in the letter if
$s$ changes in time so does both the tunneling and interactions parameters
of the EBH Hamiltonian as well as the Wannier functions themselves (an alternative approach of keeping the basis fixed in time would require a different tight basis Hamiltonian). Therefore, when $s=s(t)$ is time dependent so is the isometric basis transformation ${\cal U}(s(t))$ from the position  representation to the lattice (Wannier) representation. Of course $\mathcal{H}={\cal U}(s(t)) \mathcal{H}_X {\cal U}^\dagger(s(t)).$
Define the  transformation via $\psi(t)={\cal U}(s(t)) \psi_X(t)$ (where $\psi_X(t)$ is the wavefunction in the position representation while $\psi(t)$ corresponds to the lattice). Then a standard textbook derivation gives the proper new Hamiltonian ${\mathcal{H}_W}$ in the form:
\begin{equation}
\mathcal{H}_W  =\mathcal{H}+i\hbar\left(\frac{d}{dt} {\cal U}(s(t))\right) {\cal U}^\dagger(s(t)) = \mathcal{H}+\mathcal{W}, 
\label{Abhwdot}
\end{equation}
and the TDSE for $\psi(t): i \hbar \partial_t \psi = \mathcal{H}_W \psi.$

For $s(t)$ changing in time slowly enough, the second term may be neglected
(this corresponds physically to the assumption that the system has time to {adapt} to a given change of basis). For quick variations of $s(t)$ this term 
leads to appreciable effects, we shall now evaluate its form.

The basis for the Hilbert space for a gas of $N$ bosons consists of symmetrized (tensor) products of $N$ single particle basis functions $W_i^\alpha.$ In restricted lattice geometries, we study of multiparticle states which in transverse direction(s) contain a harmonic oscillator ground states
--- we take $W$ functions as in (\ref{eqn:bf2}) or (\ref{eqn:bf1}). These wavefunctions are effectively 1D or 2D with finite, effective ``width'',
set by the curvature of transverse confinement. We denote such a basis for $N$ particle problem by $\mathcal{W}^N(s),$ the basis
depends on the lattice height by the value of $s$ parameter through the set of single particle Wannier functions
 $\mathcal{W}(s)=\{ W_i^\alpha(\cdot,s)\}_{i,\alpha}.$

The Hilbert space for the lattice system (in which the Bose-Hubbard hamiltonian is usually expressed)
has time-independent basis (the Fock basis) $\mathcal{F}.$

Let us define a shortened notation. The lattice Fock state with occupation $n_\iota$ of mode 
$\iota$ will be denoted as $|\vec n_\mathcal{L} \rangle$
The corresponding state in the position representation for the lattice with height $s$ will be abbreviated: $|\vec n_X,s \rangle.$
We always assume that $n_1+\ldots n_\mathcal{L}=N.$ 

Action of the map ${\cal U}(s)$ from the continous space with base $\mathcal{W}^N(s)$, to the Fock space with base $\mathcal{F}$ is rather trivial: 
in the chosen orthonormal bases it is the identity matrix ---  it maps state $|\vec n_X,s(t)\rangle $ to a state $|\vec n_\mathcal{L},s(t)\rangle.$ The map is thus always an isometry
(note that for 1D and 2D lattice it is only a {\sf partial} isometry from a full continuous space).

Single particle states defined by Wannier functions and in the discrete lattice are enumerated
by two indices $\alpha$ and $i,$ from this point up to the end of the derivation, 
we introduce the multiindex $\iota=(\alpha,i)$ to simplify notation. 

Using this notation we can express the map $\cal U$ as:
$${\cal U}(s) =\sum\limits_{\vec n}  |\vec m_\mathcal{L}\rangle\langle \vec m_X, s |$$

Now we expand the derivative of the $\cal U$ isometry. We use the fact that basis of the Fock space $\mathcal{F}$ is time-independent:
$$\left(\frac{d}{dt}{\cal U}(s(t)) \right)| \psi \rangle_X  =\sum\limits_{\vec m}  |\vec m_\mathcal{L}\ \rangle \left(\frac{d}{dt}\langle \vec {m}_X, s(t)|\right) | \psi \rangle_X.$$
Thus the $\cal{W}$ term is just:
\bea
 \mathcal{W}&=&i\hbar  \sum\limits_{\vec n, \vec m}  |\vec m_\mathcal{L}\ \rangle\langle \vec n_\mathcal{L} |  \left(\frac{d}{dt}\langle \vec {m}_X, s(t)|\right) |\vec n_X, s(t)\rangle= \nonumber\\
  &=& - i\hbar \sum\limits_{\vec n, \vec m}  |\vec m_\mathcal{L}\ \rangle\langle \vec n_\mathcal{L} |  \langle \vec {m}_X, s(t)|\left(\frac{d}{dt} |\vec n_X, s(t)\rangle\right)
\label{eqn:mn}
\eea
The relation between $\vec n$ and $\vec m$ that may give nonzero contribution to the above sum remains to be
worked out as well as exact values of the coefficient. To do so, we expand the time derivative,
by inserting exact action of the symmetrization operator:
\begin{eqnarray}
{\sqrt{N!n_1!n_2!\ldots n_{\iota_0}!}} \frac{d}{dt} | {\vec n}_X, s(t) \rangle = \nonumber\\
 \sum\limits_{{\pi\in S(N)}} \frac{d}{dt}\left[W_1(x_{\pi(1)})W_1(x_{\pi(2)})\ldots \right. W_1(x_{\pi(n_1)})\cdot \nonumber\\ 
\left.\cdot W_2(x_{\pi(n_1+1)})\ldots W_2(x_{\pi({n_1+n_2})})\ldots W_{\iota_0}(x_{\pi(N)})\right] \nonumber\\
=\sum\limits_{k=1}^N\sum\limits_{{\pi\in S(N)}}\left[W_1(x_{\pi(1)})W_1(x_{\pi(2)})\ldots \right. W_1(x_{\pi(n_1)})\cdot \nonumber\\ 
\left.\cdot W_2(x_{\pi(n_1+1)})\ldots \dot{W}_\iota(x_{\pi(k)}) \ldots W_{\iota_0}(x_{\pi(N)})\right]
\label{eqn:drty}
\end{eqnarray}
In the above line each of Wannier functions $W_\iota$ depends on $t$ through $s(t).$ 
The formula is well-stated, because only finite number of modes has nonzero occupation: for
$\iota>\iota_0$ we have $n_\iota!=1$ and no factors $W_\iota.$ 

Next we use the partition of unity  $\sum |W_\varkappa \rangle\langle W_\varkappa |$,  appying it on  $\dot{W}_\iota$ we get:
\begin{equation}
\frac{d}{dt}{W}_\iota(x)= \sum\limits_{\varkappa } T^\varkappa_\iota W_\varkappa(x) 
\label{eqn:wdot2}
\end{equation}
where $T_\iota^\varkappa =  \int W_\varkappa(x) \dot{W}_\iota(x) dx.$  
Therefore, by combining together (\ref{eqn:mn}), (\ref{eqn:drty}) and (\ref{eqn:wdot2}) one obtains  that
the only $(\vec n, \vec m)$ giving nonzero contribution in (\ref{eqn:mn}) are those that
correspond to changing the mode of only one particle from configuration $\vec n$ --- the mode $\iota$ to $\varkappa.$ 
Therefore:
\be
\left\{\begin{array}{ll} m_i = n_i & i\neq \iota,\varkappa \\
m_{\iota} = n_\iota-1 & {} \\
m_{\varkappa} = n_\varkappa +1 &{}  
\end{array}
\right.
\label{eqn:relmn}
\ee
As $T_\iota^\iota=0 $ due to norm preservation, only $\iota\neq \varkappa$ terms contribute. Change of occupation is compatible with action of $a_\varkappa^\dagger a_\iota$ operator.
We will show that also the numerical factor agrees. 
A mode to be differentiated (mode $\iota$) may be chosen in (\ref{eqn:drty}) in $n_\iota$ ways, and:
\begin{equation}
\frac{n_\iota}{\sqrt{n_1!n_2!\ldots n_\mathcal{\iota_0}!}} = \frac{\sqrt{n_\iota(n_\varkappa+1)}}{{\sqrt{n_1!\ldots (n_\iota-1)!\ldots (n_\varkappa+1)!\ldots n_\mathcal{\iota_0}!}}}\\
\end{equation}
Thus from (\ref{eqn:mn}), (\ref{eqn:drty}) and (\ref{eqn:wdot2}):
\begin{eqnarray}
 \frac{d}{dt} | {\vec n}_X, s(t) \rangle=\sum\limits_{\varkappa ,\iota=1}^{\infty}T_\iota^\varkappa \sqrt{n_\iota(n_\varkappa+1)} | \vec m_X,s(t)\rangle
\label{eqn:drty2}
\end{eqnarray}
Above $\vec m$ is assumed to satisfy relations (\ref{eqn:relmn}). All in all, we obtain

\bea
\dot{\cal U}(t) {\cal U}^\dagger(t) = - \sum\limits_{\iota,\varkappa} T_\iota^\varkappa |\vec m_\mathcal{L} \rangle \langle \vec n_\mathcal{L} | \sqrt{n_\iota(n_\varkappa+1)}= \nonumber \\ 
= -\sum\limits_{\iota,\varkappa} T_{\iota}^\varkappa a_\varkappa^\dagger a_\iota.
\label{eqn:genWdot}
\eea
We now go back to the original labeling by Bloch band number: $\iota=(\alpha,i), \varkappa=(\beta,j).$ 
Now $T_\iota^\varkappa=T_{j-i}^{\beta\alpha}=-T_{i-j}^{\alpha\beta}.$
We obtain the form of $\mathcal{W}$ term used in the main article:

\begin{eqnarray}
\mathcal{W} &=& -i\hbar \sum\limits_{\iota,\varkappa} T_{i-j}^{\alpha\beta} (a_i^\alpha)^\dagger a_j^\beta,\nonumber\\  T_{i-j}^{\alpha\beta}&=&\int \dot{W}_j^\beta(x) W_i^\alpha(x) \textrm{d}^3x.
\end{eqnarray}
The term $T_\iota^\varkappa$ has to be worked out for the 
basis functions for the lattice in the appropriate dimension. In the 1D lattice, Wannier functions are of form (\ref{eqn:bf1}), then:
$T_\iota^\varkappa= \int \textrm{d}y \textrm{d}z H(y)^2 H(z)^2  \int \textrm{d}x w_j^\beta(x) \dot{w}_i^\alpha(x). $ Due to normalization:
$T_\iota^\varkappa= \int \textrm{d}x w_j^\beta(x) \dot{w}_i^\alpha(x).$
For 2D lattice, from (\ref{eqn:bf2}), we get:
$T_\iota^\varkappa= \int  \textrm{d}z H(z)^2  \int \textrm{d}x\textrm{d}y w_{j_x}^{\beta_x}(x)w_{j_y}^{\beta_y}(y) \frac{d}{dt}\left({w}_{i_x}^{\alpha_x}(x){w}_{i_y}^{\alpha_y}(y)  \right).$
Nonzero values may be obtained only if $i_x=j_x \wedge \alpha_x=\beta_x $ or $i_y=j_y \wedge \alpha_y=\beta_y.$ Thus the $\cal{W}$ term
perform hopping of a particle in only one direction (including the Bloch band change). The corresponding amplitude for hopping in $y$ direction are:
$ \int  \textrm{d}z H(z)^2 \int  \textrm{d}x w_{i_x}^{\alpha_x}(y)^2  \int\textrm{d}yw_{j_y}^{\beta_y}(y)  \dot{w}_{i_y}^{\alpha_y}(y),$ which after normalization becomes just:
$\int\textrm{d}yw_{j_y}^{\beta_y}(y)  \dot{w}_{i_y}^{\alpha_y}(y).$ Similarly for the $x$ direction we obtain:
$\int\textrm{d}xw_{j_x}^{\beta_x}(x)  \dot{w}_{i_x}^{\alpha_x}(x).$  Analogously for the 3D case (using \ref{eqn:bf3}) the amplitude for hopping in direction $t\in\{x,y,z\}$ is:
$\int\textrm{d}tw_{j_t}^{\beta_t}(t)  \dot{w}_{i_t}^{\alpha_t}(t).$

\subsection{Analysis of the revival experiment}

In the revival experiment \cite{Will2010} the atomic system is initially prepared in a relatively low lattice, on the superfluid side. Then a rapid quench is realized by a sudden increase of the lattice depth. The subsequent evolution of the system in a deep lattice is monitored by
measuring the time dependence of the contrast of interference fringes in the momentum distribution. The initial coherent state like occupation of sites evolves differently in the deep insulating-type lattice showing the decay and partial revivals of coherence. In the simulation the evolution of coherence  may be monitored by the time dependence of the order parameter $\phi=\langle a^0 \rangle$ where the superscript indicates that the lowest Bloch band is taken into account only (we drop the site index, as we shall consider a single site only - in a very deep lattice the tunneling may be neglected
and a single site evolution is considered - see \cite{Will2010}).

\begin{figure}[ht]
\centering
\includegraphics[clip,width=8.5cm]{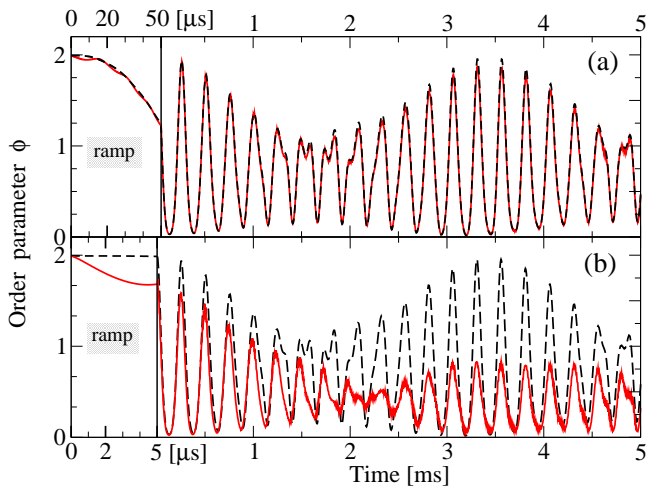}
\caption{ (color online) Time dependence of the order parameter in the lowest band, 
$\phi= \langle\psi(t)|a^0|\psi(t)\rangle$ for $50 \mu s$ initial linear quench 
as used in the experiment \cite{Will2010} (top panel) and for a faster quench 
of $5 \mu s$ (bottom panel). Red full line - exact evolution using $H_W$
Hamiltonian, Eq.(\ref{Abhwdot}), black (dashed) line the result obtained neglecting ${\cal W}$ term. The difference between the two is negligible for the top panel, showing that the higher bands play a minor role in the experiment. For the faster quench (bottom panel) the ${\cal W}$ term describing time-dependence of Wannier functions is crucial for the description of dynamics. Note that the time scale during the quench is expanded in comparison to the subsequent 
evolution to make the plot more readable.
}
\label{fig:reviv}
\end{figure}

The contribution of ${\cal W}$ term (see the letter) depends on the speed of the quench, it becomes important when excitations
of higher Bloch bands become appreciable.
The authors \cite{Will2010} wanted to avoid population of excited bands (that could be controlled experimentally) thus 
 they experimentally chose the duration of the quench, $\tau=50 \mu s$, sufficiently long so that the higher bands excitations were negligible. 
Consequently a contribution of Wigner functions dynamics (the ${\cal W}$ term) is for experimental parameters quite small (see the upper panel in Fig.~\ref{fig:reviv}. As it turns out already for 
$\tau=20\mu s$ effects due to Wannier function dynamics are appreciable
while a still faster  
 $5 \mu s$ quench  leads to  strong Wannier functions dynamics effects (compare Fig.~\ref{fig:reviv}b). Let us point out that to make calculations less computer demanding we used a 2D lattice,
the initial state, $\psi(0)$ was prepared as a coherent state with $\langle n_0\rangle=2$ at $s_0=8$ and a linear quench up to $s_1=40$ was realized. That roughly corresponds to the revival plot in Fig.~2 of \cite{Will2010}. 

It seems, therefore, that making the quench time in the experiment \cite{Will2010} shorter by an order of magnitude would allow for a direct experimental verification of the approach presented here.


\end{document}